\documentclass[11pt]{iopart}

\usepackage[german,english]{babel}
\usepackage{amssymb}
\usepackage{amsfonts}
\usepackage{mathrsfs}
\usepackage{epsfig}

\begin{document}

\title[Blinking statistics of a molecular beacon]{Blinking statistics of a
molecular beacon triggered by end-denaturation of DNA}

\author{Tobias Ambj\"ornsson and Ralf Metzler}
\address{NORDITA -- Nordic Institute for Theoretical Physics,\\ 
Blegdamsvej 17, DK-2100 Copenhagen \O, Denmark}
\ead{\mailto{ambjorn@nordita.dk},\mailto{metz@nordita.dk}}

\begin{abstract}
We use a master equation approach based on the Poland-Scheraga free energy
for DNA denaturation to investigate the (un)zipping dynamics of a denaturation
wedge in a stretch of DNA, that is clamped at one end. In particular, we
quantify the blinking dynamics of a fluorophore-quencher pair mounted within
the denaturation wedge. We also study the behavioural changes in the presence
of proteins, that selectively bind to single-stranded DNA. We show that such
a setup could be well-suited as an easy-to-implement nanodevice for sensing
environmental conditions in small volumes.
\end{abstract}


\section{Introduction}

During the last decade or so, technical progress in detection and
manipulation of single molecules and their dynamics has snowballed. By
fluorescence spectroscopy, optical tweezers, atomic force microscopy,
or patch clamp techniques, for instance, it is possible on the single
molecule level to probe the opening and closing dynamics of local
denaturation zones in a DNA molecule \cite{oleg}, to study the binding
of single-stranded DNA binding proteins to overstretched DNA
\cite{pant,pant1}, to stepwise disrupt domains in a protein \cite{rief}, or
to monitor the passage of a biopolymer through a nanopore in a
membrane \cite{Kasianowicz96,amitrev}, just to name a few. This
experimental progress is accompanied by advances in the theoretical
understanding of fundamental processes relevant on small scales such
as the Jarzinsky relation connecting measurements of the
nonequilibrium work needed, e.g., to stretch an RNA segment
\cite{liphardt}, to the difference in the corresponding thermodynamic
potential \cite{jarzinsky}; or the entropy production along single
trajectories exposed to stochastic forces \cite{seifert}. Given the
novel possibilities to synthesize supramolecules with topologically
confined mechanical units \cite{lehn}, entropy-based designer
molecules such as molecular muscles were proposed \cite{cpl,angew},
and new possibilities to produce dynamic nanosensors discussed
\cite{ctn}.

\begin{figure}
\begin{center}
\scalebox{0.5}{\epsfbox{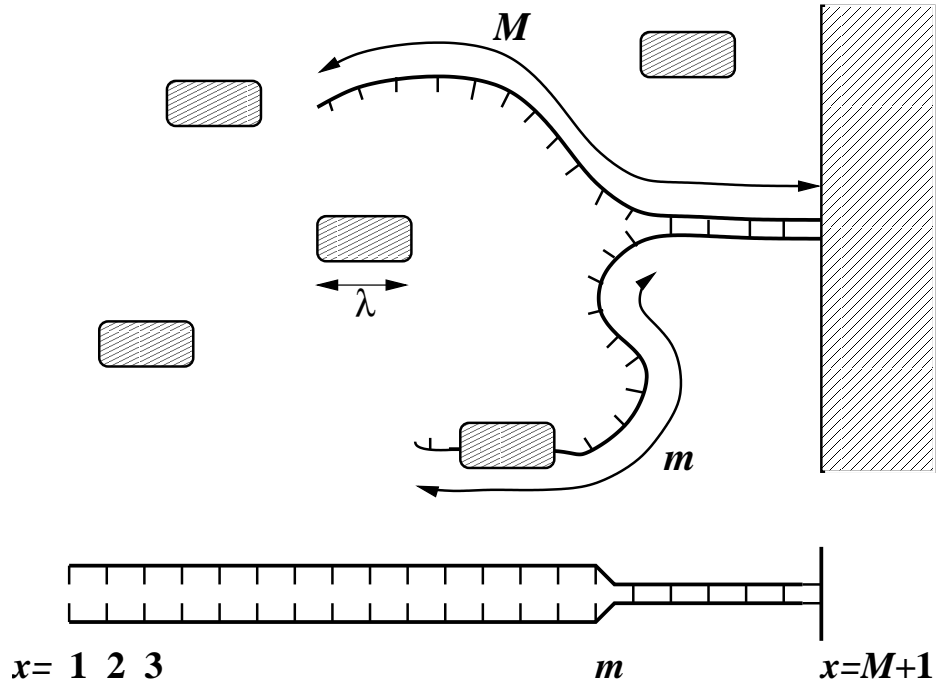}}
\end{center}
\caption{Schematic of the end-denaturing of a double-stranded DNA molecule,
that is clamped at one end; here, by attachment to a wall. The number of
denatured base-pairs is $m$, and the overall length of the DNA segment
is $M$. Selectively single-stranded DNA binding proteins from the
surrounding solution bind to the denatured portion of the DNA. Once bound,
the SSBs prevent closing of the denaturation wedge. In reality, the clamping
can be achieved by sealing the denaturation zone created by AT base-pairs
with a stretch of more stable GC base-pairs, compare reference \cite{oleg}.}
\label{scheme}
\end{figure}

In what follows, we explore the dynamics of a stretch of
double-stranded DNA (dsDNA), that is clamped at one end but allowed to
denature from the other, as sketched in Figure \ref{scheme}. Internal bubble
formation, in comparison, is suppressed by a Boltzmann factor $\sigma_0\simeq
10^{-5}\ldots 10^{-3}$ \cite{blake}, and this effect can therefore be
neglected, see below. In the setup we have in mind an individual base-pair is
tagged with a fluorophore-quencher pair, see Figure \ref{beacon}: Once
separated, the dye starts to fluoresce, and the resulting blinking can
be detected \cite{olegrev}. Our aim is to establish a quantitative description
of such a \emph{molecular beacon}. We will use a master equation approach to
describe the sequential unzipping and zipping dynamics of the
denaturation wedge, also taking into account the possible presence of
proteins, that selectively bind to single-stranded DNA (ssDNA). These
single-stranded DNA binding proteins (SSBs) occur in the cells of most
organisms.
Here, we show that the autocorrelation function for the blinking
depends on the concentration of SSBs and the binding strength between
the single strand and the SSBs (the latter varies, for instance, with
salt concentration) as well as temperature.  The denaturation wedge
depicted in figure \ref{beacon}
therefore acts as a nanosensor that can be probed on the single
molecule level using fluorescence techniques.

\section{Experimental setup: Quantifying the blinking dynamics of the
molecular beacon}

\begin{figure}
\begin{center}
\scalebox{0.5}{\epsfbox{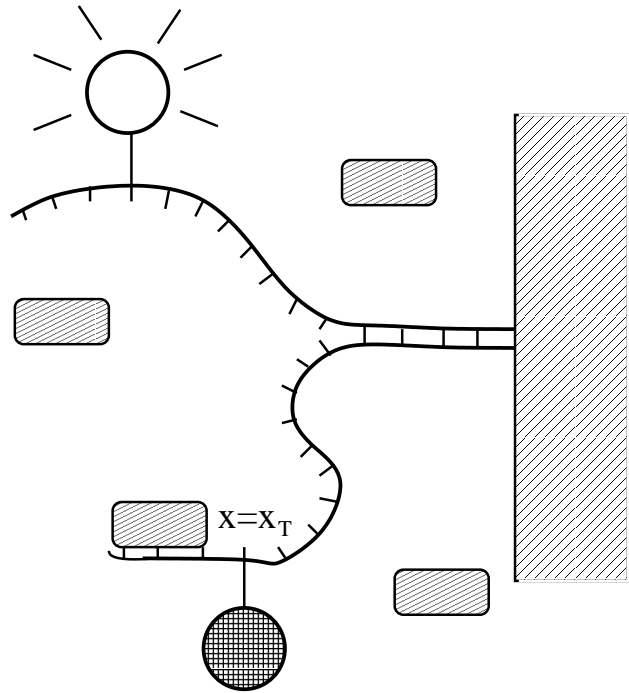}}
\end{center}
\caption{Schematic of the molecular beacon setup. A fluorophore starts to
fluoresce in the incident laser light, once the denaturation wedge moves
the fluorophore apart from the quencher.}
\label{beacon}
\end{figure}

Before setting out to describe our general theoretical scheme, we
first describe the experimental setup we have in mind in some
detail. We consider modelling the blinking behaviour of a
fluorophore-quencher pair mounted on the denaturation wedge as shown
in Figure \ref{beacon}. This setup, similar to the ones described in
References \cite{oleg,olegrev} works as follows.
As long as the dsDNA is intact, fluorophore and quencher are in close
proximity. Once they come apart from one another when the denaturation
wedge opens up, the incident laser light causes fluorescence of the dye.
The on/off blinking of this "molecular beacon" can be monitored in the
focus of a confocal microscope.

The blinking renders immediate information about the state of the base-pair,
that is tagged by the dye-quencher pair. Blinking, that is, indicates that
the base-pair is currently broken. It is therefore advantageous to define
the random variable $I(t)$ with the property
\begin{equation}
I(t)=\left\{\begin{array}{ll}
0 & \mbox{if base-pair at $x=x_T$ is closed}\\
1 & \mbox{if base-pair at $x=x_T$ is open}\end{array}\right.,
\end{equation}
and we are interested in the behaviour of the autocorrelation function
\begin{equation}
A(t)=\langle I(t)I(0)\rangle-\langle I\rangle_{\mathrm{eq}^2},\label{A_t}
\end{equation}
where $\langle I\rangle_{\mathrm{eq}}$ is the (ensemble) equilibrium value.
Given the fact that the formation of an internal denaturation bubble
is connected with a rather high initiation barrier $\sigma_0\simeq 10^{-5}
\ldots 10^{-3}$ (corresponding to $7\ldots 12$ $k_BT$ at room temperature),
such bubbles are much less probable than denaturation from the unclamped end,
we focus only on the end-denaturation. Therefore,
a base-pair at $x=x_T$ is open if $m\ge x_T$, see figure \ref{beacon}. A
word on the interpretation of the average $\langle I(t)I(0)\rangle$ is
in order. Denoting by $\rho(I,t;I',0)$ the associated joint
probability density that the tagged base-pair is in state $I$ at time
$t$ given that it was in state $I'$ at initial time $t=0$, we can
rewrite the autocorrelation function as
\begin{equation}
\langle I(t)I(0)\rangle=\sum_{I,I'}I\rho(I,t;I',0)I'=\rho(1,t;1,0).
\end{equation}
This is but the survival probability density for a denaturation wedge, i.e.,
the probability density that the base-pair is open at time $t$, given that is
was open at system preparation at time $t=0$. 

In the remaining part of this study we present a general scheme for
calculating the opening-and closing dynamics for the physical system
presented in figures \ref{scheme} and \ref{beacon}. The focus is on
presenting a scheme which allows for the calculation of the
measurable quantity $A(t)$ defined above. We note, however, that our
scheme is sufficiently general to allow for straightforward derivation
of other experimental quantities, such as the binding dynamics of SSBs in
optical tweezers overstretching setups \cite{pant,pant1,igor}. In that case,
the Boltzmann factor for opening a base-pair, $u=\exp(\beta\epsilon)$
($\epsilon$ being the binding energy of a base-pair, see below) becomes
modified to $u=\exp\left(\beta\epsilon-\mathcal{T}\theta_0\right)$, where
$\mathcal{T}$ is the external torque exerted by the optical tweezers setup,
and $\theta_0=2\pi/10.35$ is the relaxed dsDNA-twist per base-pair.

\section{Master equation for end-unzipping}

\begin{figure}
\begin{center}
\scalebox{0.4}{\epsfbox{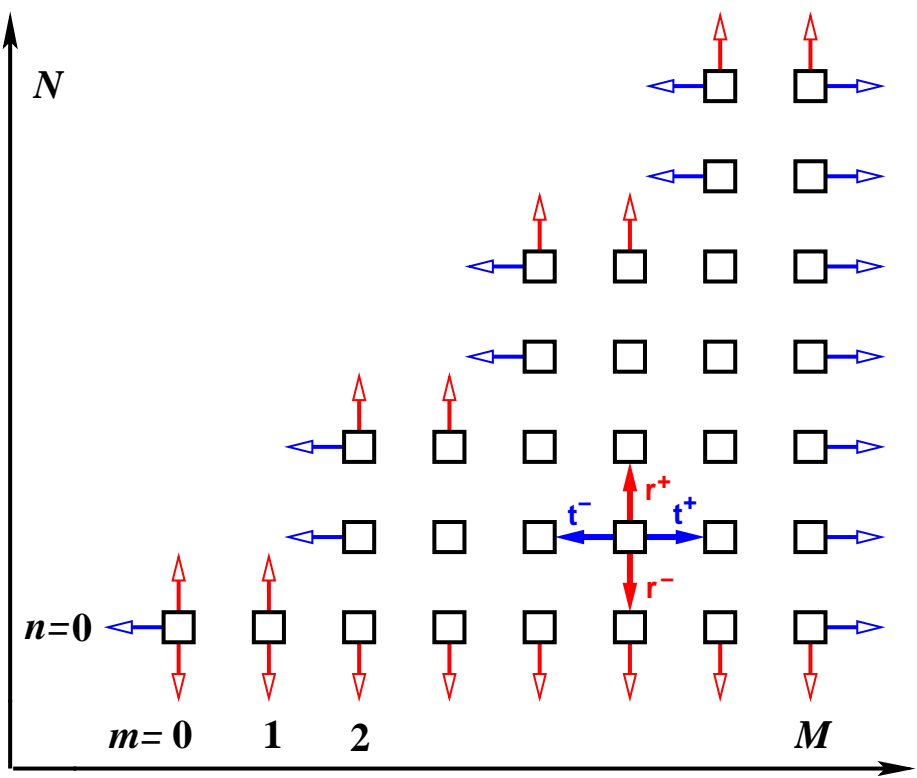}}
\end{center}
\caption{Configuration lattice showing the possible configurations
$\Box$ of the system in the $(m,n)$-plane. The empty arrow heads
represent forbidden moves due to the boundary conditions. The maximum
number of bound SSBs is $N=2[M/\lambda]$. }
\label{lattice}
\end{figure}

Denote by $P(m,n,t)$ the probability distribution that there are $m$ broken
base-pairs and $n$ bound SSBs at time $t$. As $m$ and $n$ are the slow
variables of the system, their dynamics can be described in terms of the
master equation
\begin{eqnarray}
\nonumber
\frac{\partial P(m,n,t)}{\partial t}&=&\mathsf{t}^+(m-1,n)P(m-1,n,t)+
\mathsf{t}^-(m+1,n,t)P(m+1,n,t)\\
\nonumber
&&\hspace{0.8cm}-\left[\mathsf{t}^+(m,n)+\mathsf{t}^-(m,n)\right]P(m,n,t)\\
\nonumber
&&+\mathsf{r}^+(m,n-1)P(m,n-1)+\mathsf{r}^-(m,n+1)P(m,n+1,t)\\
&&\hspace{0.8cm}-\left[\mathsf{r}^+(m,n)+\mathsf{r}^-(m,n)\right]P(m,n,t),
\label{master}
\end{eqnarray}
compare the discussion in References \cite{ssb1,ssb,ambme}. Mediated by the
transfer rates $\mathsf{t}^\pm$ and $\mathsf{r}^\pm$, the population of a given
state $(m,n)$ is continuously changed by (un)zipping a further base-pair
and (un)binding of an SSB. To complete the master equation (\ref{master}),
we need to specify the boundary conditions: The clamping to the right at
base-pair $x=M+1$ ensures that no further unzipping beyond base-pair $M$
occurs, namely,
\begin{equation}
\label{b1}
\mathsf{t}^+(M,n)=0.
\end{equation}
Moreover, if both branches of the denaturation wedge are fully occupied by
SSBs, i.e., the maximum number of SSBs,
\begin{equation}
n^{\mathrm{max}}(m)=2\left[\frac{m}{\lambda}\right],
\end{equation}
is bound, then the base-pair at the zipping fork is not allowed to close:
\begin{equation}
\label{b2}
\mathsf{t}^-\left(m=\frac{n\lambda}{2},n=n^{\mathrm{max}}(m)\right)=0.
\end{equation}
Also, if only one of the two branches of the wedge is fully occupied
the zipper cannot close:
\begin{equation}
\label{b3}
\mathsf{t}^-\left(m=\frac{(n+1)\lambda}{2},n=n^{\mathrm{max}}(m)-1\right)=0.
\end{equation}
Similar to the (un)zipping rates $\mathsf{t}^{\pm}$, we impose the
boundary condition
\begin{equation}
\mathsf{r}^+\left(m,n^{\mathrm{max}}(m)\right)=0,
\end{equation}
i.e., once the denaturation wedge is completely occupied, no additional
SSB is allowed to bind; and when $n=0$ SSBs are bound, no further SSB
can detach:
\begin{equation}
\mathsf{r}^-\left(m,0\right)=0,
\end{equation}
The configuration lattice showing the allowed moves is illustrated in
Figure \ref{lattice}. Empty arrow heads indicate forbidden moves.

The general solution of the master equation (\ref{master}) can be obtained
through the ansatz
\begin{equation}
P(m,n,t)=\sum_pc_pQ_p(m,n)e^{-\eta_pt},
\end{equation}
corresponding to an expansion in eigenmodes. Here, the expansion coefficients
$c_p$ are determined by the initial condition. Inserting above expansion into
equation (\ref{master}) produces the eigenvalue equation
\begin{eqnarray}
\nonumber
-\eta_pQ_p(m,n)&=&\mathsf{t}^+(m-1,n)Q_p(m-1,n)+\mathsf{t}^-(m+1,n)Q_p(m+1,n)\\
\nonumber
&&\hspace{0.8cm}-\left[\mathsf{t}^+(m,n)+\mathsf{t}^-(m,n)\right]Q_p(m,n)\\
\nonumber
&&+\mathsf{r}^+(m,n-1)Q_p(m,n-1)+\mathsf{r}^-(m,n+1)Q_p(m,n+1)\\
&&\hspace{0.8cm}-\left[\mathsf{r}^+(m,n)+\mathsf{r}^-(m,n)\right]Q_p(m,n)
\label{eigen1}
\end{eqnarray}
for the $p$th eigenmode. The concrete forms for the transfer rates are
defined in the next section.

\section{Partition function and transfer rates}

In order to obtain the transfer coefficients $\mathsf{t}^{\pm}$ and
$\mathsf{r}^{\pm}$, we first define the partition function
$\mathscr{Z}(m,n)$ for the end-denaturation fork of the dsDNA and the
SSBs bound to its two branches. To this end, we note that we can
decouple the partition coefficient
\begin{equation}
\mathscr{Z}(m,n)=\mathscr{Z}^{\mathrm{DNA}}(m)\mathscr{Z}^{\mathrm{SSB}}(m,n)
\end{equation}
into the contributions $\mathscr{Z}^{\mathrm{DNA}}$ counting the degrees of
freedom of the DNA molecule, and the contribution $\mathscr{Z}^{\mathrm{SSB}}$
of the SSBs. According to the Poland-Scheraga model of DNA denaturation, we
have
\begin{equation}
\mathscr{Z}^{\mathrm{DNA}}(m)=u^m.
\end{equation}
Here, $u=\exp\left(\beta\epsilon\right)$ is the Boltzmann factor for
opening a base-pair, i.e., the activation needed to overcome the free
energy barrier $\epsilon$ for opening an additional
base-pair. $\epsilon$ combines two appreciable contributions from
enthalpy cost and entropy gain on breaking the base-pair, that almost
cancel such that $\epsilon$ is of the order of a $k_BT$ at
physiological temperature and salt: At 37$^{\circ}$, $u\approx 0.6$ in
a zone of AT base-pairs, and $u\approx 0.2$ in a GC-domain
\cite{blake}. We note that the parameter $u$ is sensitive to salt
concentration. Here, we consider homopolymer zones of either AT
or GC; for the treatment of a heteropolymer, see Reference
\cite{hetero}. As already mentioned, we also neglect the formation of internal
denaturation bubbles within the dsDNA stretch, that would require the crossing
of the initiation barrier $\sigma_0$. This contrasts our previous
studies \cite{hame,ssb1,ssb,hetero,suman}, in which we focused on
internal bubbles, preventing end-denaturation wedges by clamping at
both ends. Note also that in the case of the end-wedges, we do not
have to take care of the entropy loss due to loop formation as present
in internal denaturation bubbles.

The contribution from the SSBs has the form
\begin{equation}
\mathscr{Z}^{\mathrm{SSB}}(m,n)=\kappa^n\Omega^{\mathrm{SSB}}(m,n),
\end{equation}
sharing the energetic component
\begin{equation}
\kappa=c_0K^{\mathrm{eq}}
\end{equation}
per bound SSB, where $c_0$ is the SSB concentration in solution and
$K^{\mathrm{eq}}=v_0\exp\left(|E_{\mathrm{bind}}|/k_BT\right)$ the
equilibrium binding constant of SSB-binding, with the typical volume $v_0$ of
the SSBs and the binding energy $E_{\mathrm{bind}}$;
and the number of possible ways of arranging $n$ SSBs on the two arches of
the denaturation fork, both being of length $m$ \cite{ssb1,ssb}:
\begin{equation}
\Omega^{\mathrm{SSB}}(m,n)=\left.\sum_{n'=0}^n\omega^{\mathrm{SSB}}(m,n')
\omega^{\mathrm{SSB}}(m,n-n')\right|^{n-n'\le n^{\mathrm{max}}/2}
_{n'\le n^{\mathrm{max}}/2}.
\end{equation}
This counting allows to have a different number of SSBs on the two arches.
The degrees of freedom of distributing $n$ SSBs each covering $\lambda$ bases,
on an arch of size $m$ is
\begin{equation}
\omega^{\mathrm{SSB}}(m,n)={m-(\lambda-1)n \choose n}=\frac{(m-[\lambda-1]n)!}{
n!(m-\lambda n)!}.
\end{equation}

From the partition function, we can now come to specify the transfer rates.
As a first requirement we impose that the rates obey the detailed balance
condition \cite{vankampen}
\begin{equation}
\mathsf{r}^+(m,n-1)\mathscr{Z}(m,n-1)=\mathsf{r}^-(m,n)\mathscr{Z}(m,n)
\end{equation}
for SSB (un)binding, and
\begin{equation}
\mathsf{t}^+(m-1,n)\mathscr{Z}(m-1,n)=\mathsf{t}^-(m,n)\mathscr{Z}(m,n)
\end{equation}
for base-pair (un)zipping. Detailed balance guarantees that the probability
distribution $P(m,n,t)$ for long times relaxes towards the Boltzmann
distribution. However, the detailed balance condition does not uniquely
specify the rates. Using the remaining freedom of choice \cite{vankampen,ssb1},
we settle for the following forms. In the case of SSB (un)binding, we choose
\begin{equation}
\mathsf{r}^+(m,n)=(n+1)q\frac{\mathscr{Z}(m,n+1)}{\mathscr{Z}(m,n)}=
(n+1)q\kappa\frac{\Omega^{\mathrm{SSB}}(m,n+1)}{\Omega^{\mathrm{SSB}}(m,n)}
\end{equation}
for the binding rate, and
\begin{equation}
\mathsf{r}^-(m,n)=nq
\end{equation}
for unbinding. The unbinding rate, that is, is proportional to the number
of bound SSBs, as it should. Apart from this $n$-factor, we choose it to
solely depend on the constant unbinding rate $q$. Conversely, the binding
rate includes, apart from this rate $q$, the binding strength $\kappa$,
the relative number of degrees of freedom given by the ratio of the
$\Omega^{\mathrm{SSB}}$ factors; and the factor $(n+1)$. The factor
$(n+1)\Omega^{\mathrm{SSB}}(m,n+1)/\Omega^{\mathrm{SSB}}(m,n)$ accounts
for the combinatorial number of ways to put an additional SSB of size
$\lambda$ onto either arch of the denaturation fork. In the limit
$\lambda=1$ the resulting expression for $\mathsf{r}^+(m,n)$
becomes $q\kappa(m-n)$ and thus designates the number of free binding
sites, compare the discussion in \cite{ssb1,vankampen}.

Similarly, we choose a completely asymmetric form for the base-pair
(un)zipping rates. Thus, for unzipping of an additional base-pair we
use
\footnote{We here, for simplicity, neglect any $m$-dependence of the
rate constant $k$, compare the discussion about the 'hook'-effect in
references \cite{ssb1,ssb}.}
\begin{equation}
\mathsf{t}^+(m,n)=\mathsf{t}^+(m)=k\frac{\mathscr{Z}^{\mathrm{DNA}}(m+1,n)}{
\mathscr{Z}^{\mathrm{DNA}}(m,n)}=ku,
\end{equation}
carrying the full Boltzmann factor $u$, apart from the rate constant $k$. The
zipping rate
\begin{equation}
\mathsf{t}^-(m,n)=k\frac{\mathscr{Z}^{\mathrm{DNA}}(m-1,n)}{\mathscr{Z}^{
\mathrm{DNA}}(m,n)}=k\frac{\Omega^{\mathrm{SSB}}(m-1,n)}{\Omega^{\mathrm{SSB}}
(m,n)}
\end{equation}
contains all information related to the interplay with the number $m$ of
bound SSBs, and is proportional to the probability that the base-pair next
to the denaturation fork is unoccupied.
This choice realistically describes the fact that a region
almost fully occupied with SSBs is less likely to decrease in size. We
note that in general $\mathsf{t}^-(m,n)\le k$, and that $\mathsf{t}^-(m,0)
=k$. Note also that we can conveniently define a dimensionless parameter
\begin{equation}
\gamma\equiv\frac{q}{k}
\end{equation}
that measures the competition between SSB (un)binding and base-pair
(un)zipping.

\section{Results: Blinking autocorrelation of the molecular beacon}

Having defined the dynamics of the end-denaturation wedge in terms of
the numbers of broken base-pairs, $m$, and of bound SSBs, $n$, we now
proceed to calculate the autocorrelation $A(t)$ for the blinking
behaviour of a fluorophore-quencher pair mounted on the denaturation
wedge as shown in Figure \ref{beacon}. Following the results from reference
\cite{hetero}, we can express $A(t)$ according to equation \eref{A_t} in
the form
\begin{equation}
A(t)=\sum_{p\neq 0}T_p^2e^{-t/\tau_p},\label{eq:A_t}
\end{equation}
with relaxation times $\tau_p\equiv\eta_p^{-1}$, and where
\begin{equation}
T_p=\left.\sum_{m,n}Q_p(m,n)\right|_{m\ge x_T}.\label{eq:T_p}
\end{equation}
It should also be noted that, in order to be detected experimentally,
it might be necessary that the fluorophore-quencher pair should be separated
further than by the separation provided by solely opening the base-pair $x_T$.
Thus, if $\Delta$ additional base-pairs need to be broken
before fluorescence occurs, the quantity $T_p$ will be modified to
$T_p=\left.\sum_{m,n}Q_p(m,n)\right|_{m\ge x_T+\Delta}$.

Alternatively, we can rewrite the autocorrelation function $A(t)$ according
to the spectral decomposition
\begin{equation}
A(t)=\int_0^{\infty}f(\tau)\exp\left(-\frac{t}{\tau}\right)d\tau.
\end{equation}
This way, we obtain the weighted spectral density ("relaxation time
spectrum")
\begin{equation}
f(\tau)=\sum_{p\neq 0}T_p^2\delta\left(\tau-\tau_p\right).
\end{equation}
This distribution $f(\tau)$ renders information about the contributions of
individual relaxation modes to the autocorrelation function $A(t)$, see also
below.

For the two limiting cases corresponding to absence of SSBs, and to
fast (un)binding dynamics of the SSBs, we obtain analytical solutions for
$A(t)$ from the master equation (\ref{master}) below. The general case is
solved numerically along similar lines as developed in references \cite{ssb1}.

\subsection{Absence of SSBs}

If no SSBs are present, the (un)zipping rates $\mathsf{t}^{\pm}$ assume the
simpler form
\begin{equation}
\overline{\mathsf{t}}^+(m)=\mathsf{t}^+(m,0)=ku,
\end{equation}
and
\begin{equation}
\overline{\mathsf{t}}^-(m)=\mathsf{t}^+(m,0)=k,
\end{equation}
corresponding the transfer rates for an asymmetric random walk. This
contrasts the non-linear form for internal bubbles that includes the loop
entropy loss \cite{ssb1,ssb}. The master equation reduces to the one-variable
form
\begin{eqnarray}
\nonumber
\frac{\partial \overline{P}(m,t)}{\partial t}&=&\overline{\mathsf{t}}^+(m-1)
\overline{P}(m-1,t)+\overline{\mathsf{t}}^-(m+1)\overline{P}(m+1,t)\\
&&-\left[\overline{\mathsf{t}}^+(m)+\overline{\mathsf{t}}^-(m)\right]
\overline{P}(m,t)
\label{master1}
\end{eqnarray}
with the eigenvalue decomposition
$\overline{P}(m,t)=\sum_p\overline{c}_p\overline{Q}_p(m)e^{-\overline{\eta}
_pt}$. Thus, we obtain the eigenvalue equation
\begin{eqnarray}
\nonumber
-\overline{\eta}_p\overline{Q}_p(m)&=&\overline{\mathsf{t}}^+(m-1)
\overline{Q}_p(m-1)+\overline{\mathsf{t}}^-(m+1)\overline{Q}_p(m+1)\\
&&-\left[\overline{\mathsf{t}}^+(m)+\overline{\mathsf{t}}^-(m)\right]
\overline{Q}_p(m),
\end{eqnarray}
with the obvious boundary conditions
$\overline{\mathsf{t}}^-(0)=0$, and $\overline{\mathsf{t}}^+(M)=0$.
The solution of the reduced eigenvalue equation for $\overline{Q}_p$ can
be obtained in terms of orthogonal polynomials (or Chebyshev polynomials)
according to References \cite{ssb1,ledermann}:
\begin{equation}
\overline{Q}_p(m)=\frac{u^{m/2}}{\sin \omega_p}\left\{\sin\left[(m+1)\omega_p
\right]-u^{-1/2}\sin\left[m\omega_p\right]\right\},
\end{equation}
where the eigenvalues become
\begin{equation}
\overline{\eta}_p=k\left[u+1-2u^{1/2}\cos\omega_p\right]
\end{equation}
with
\begin{equation}
\omega_p=\frac{p\pi}{M+1}.
\end{equation}
We notice that the relaxation times $\bar{\tau}_p\equiv 1/\bar{\eta}_p$
fulfil the inequalities
\begin{equation}
\tau_{\mathrm{min}}(u)\equiv k^{-1}\left(1+u^{1/2}\right)^{-2}\le
\overline{\tau}_p \le k^{-1}\left(1-u^{1/2}\right)^{-2}\equiv\tau_{
\mathrm{max}}(u).
\end{equation}
Thus, at the melting transition where $u=1$, and for an infinitely
long DNA segment $M\rightarrow \infty$ the longest relaxation time
diverges, i.e., the denaturation wedge has a diverging lifetime, as
one would expect. The correlation function $A(t)$ corresponding to the
results above are shown by the black dash-dotted
curves in figure \ref{fig:A_t}.

We note that conditions similar to the absence of SSBs are fulfilled if the
concentration of SSBs is very low, $\kappa\to 0$, or if $\gamma\kappa\to 0$,
compare the existence of a kinetic block to SSB-binding in DNA-bubbles
\cite{pant,pant1,ssb1,ssb}.

\subsection{Fast (un)binding dynamics of SSBs}

In case the (un)binding dynamics of the SSBs is much faster than the typical
base-pair (un)zipping rates, $\gamma\gg 1$, we can adiabatically eliminate the
degrees of freedom corresponding to the SSB dynamics \cite{risken}. Ensuing are
an effective free energy landscape, dressed by the SSB-interactions with the
denaturation wedge, and a reduced one-variable master equation of the type
(\ref{master1}), but with the following (dressed) rate coefficients \cite{ssb1}:
\begin{equation}
\widetilde{\mathsf{t}}^\pm(m)=\sum_{n=0}^{n^{\mathrm{max}}(m)}\mathsf{t}^+(m,n)
\frac{\mathscr{Z}(m,n)}{\mathscr{Z}(m)},
\end{equation}
where $\mathscr{Z}(m)\equiv\sum_n\mathscr{Z}(m,n)$.  When calculating
the two dressed rates $\widetilde{\mathsf{t}}^{\pm}$, it is important
to consider the boundary conditions (\ref{b1}), (\ref{b2}), and
(\ref{b3}). From these we also deduce the modified boundary conditions
$\widetilde{\mathsf{t}}^+(M)=0$ and $\widetilde{\mathsf{t}}^-(0)=0$.
The corresponding result for the autocorrelation function are shown by
the dashed curves in figure \ref{fig:A_t}.

\subsection{General case}

\begin{figure}
\begin{center}
\scalebox{0.42}{\epsfbox{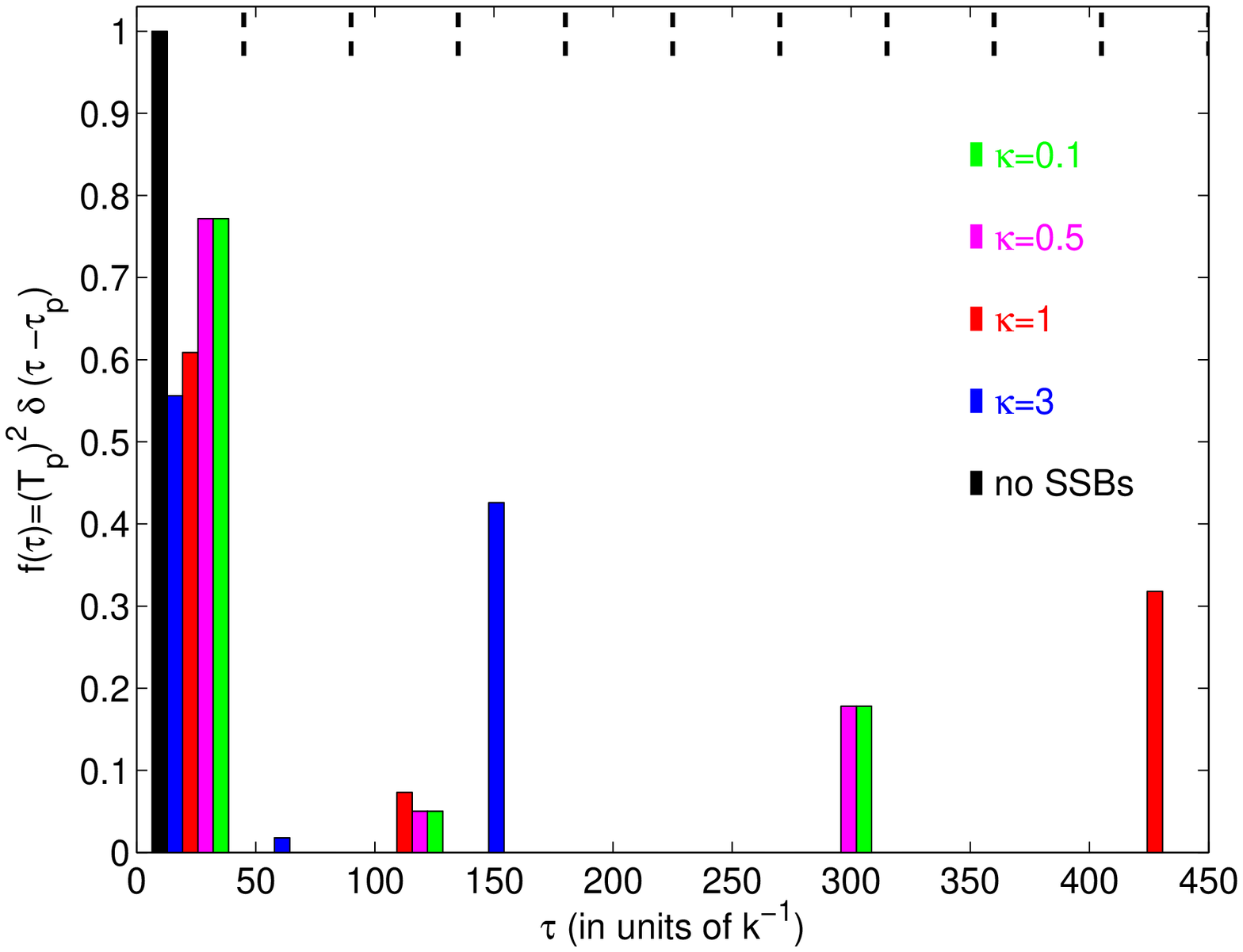}}
\scalebox{0.42}{\epsfbox{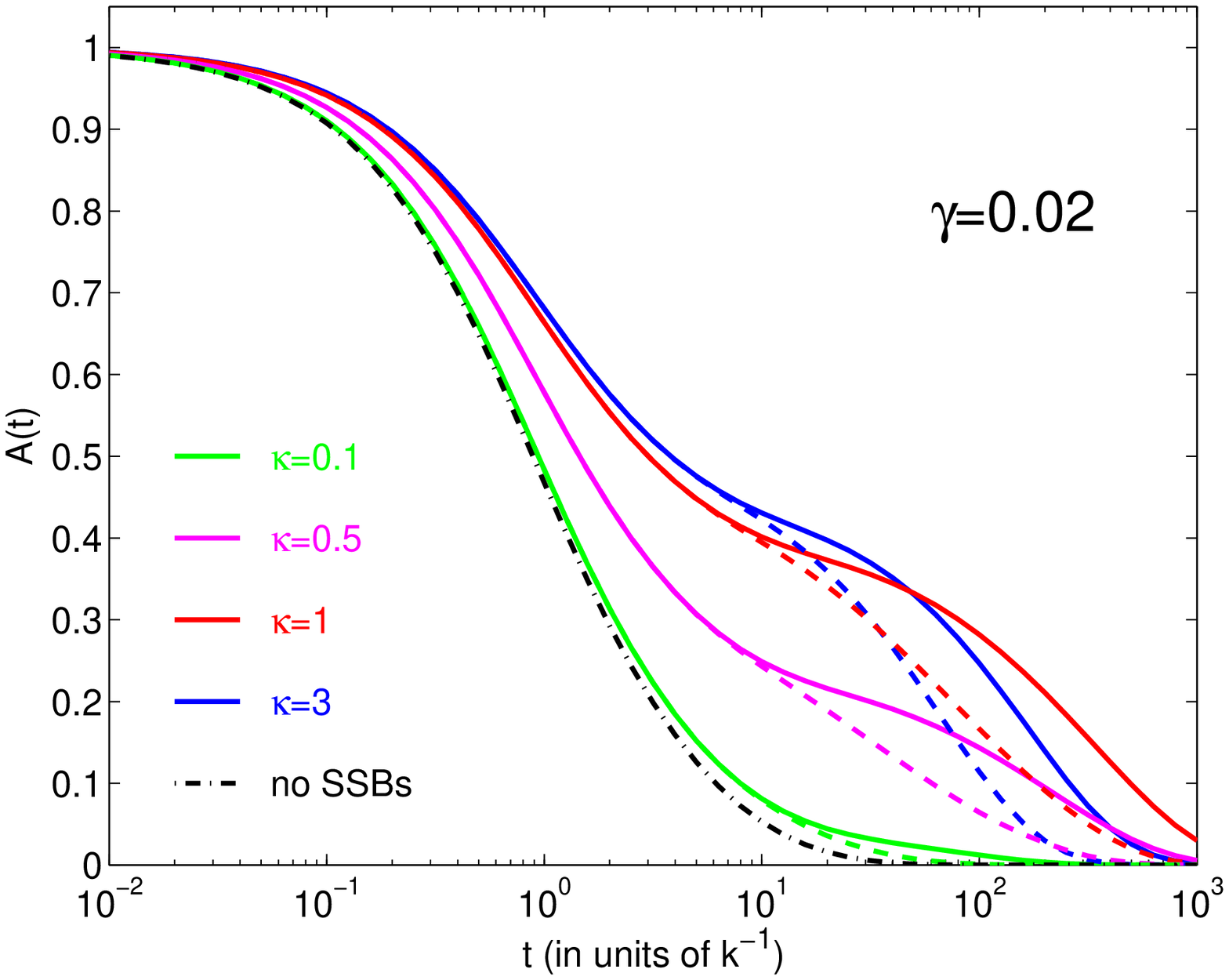}}
\end{center}
\begin{center}
\scalebox{0.42}{\epsfbox{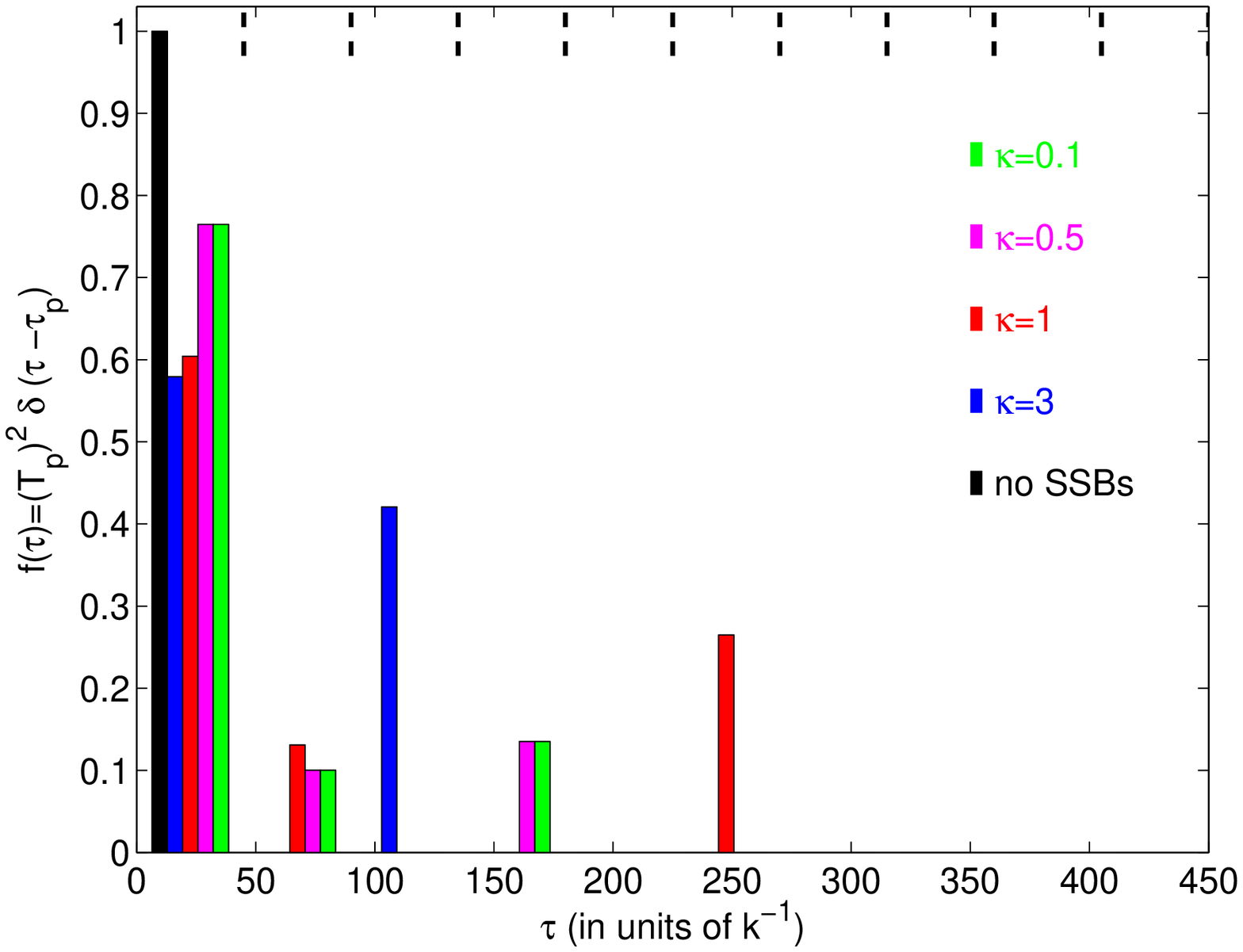}}
\scalebox{0.42}{\epsfbox{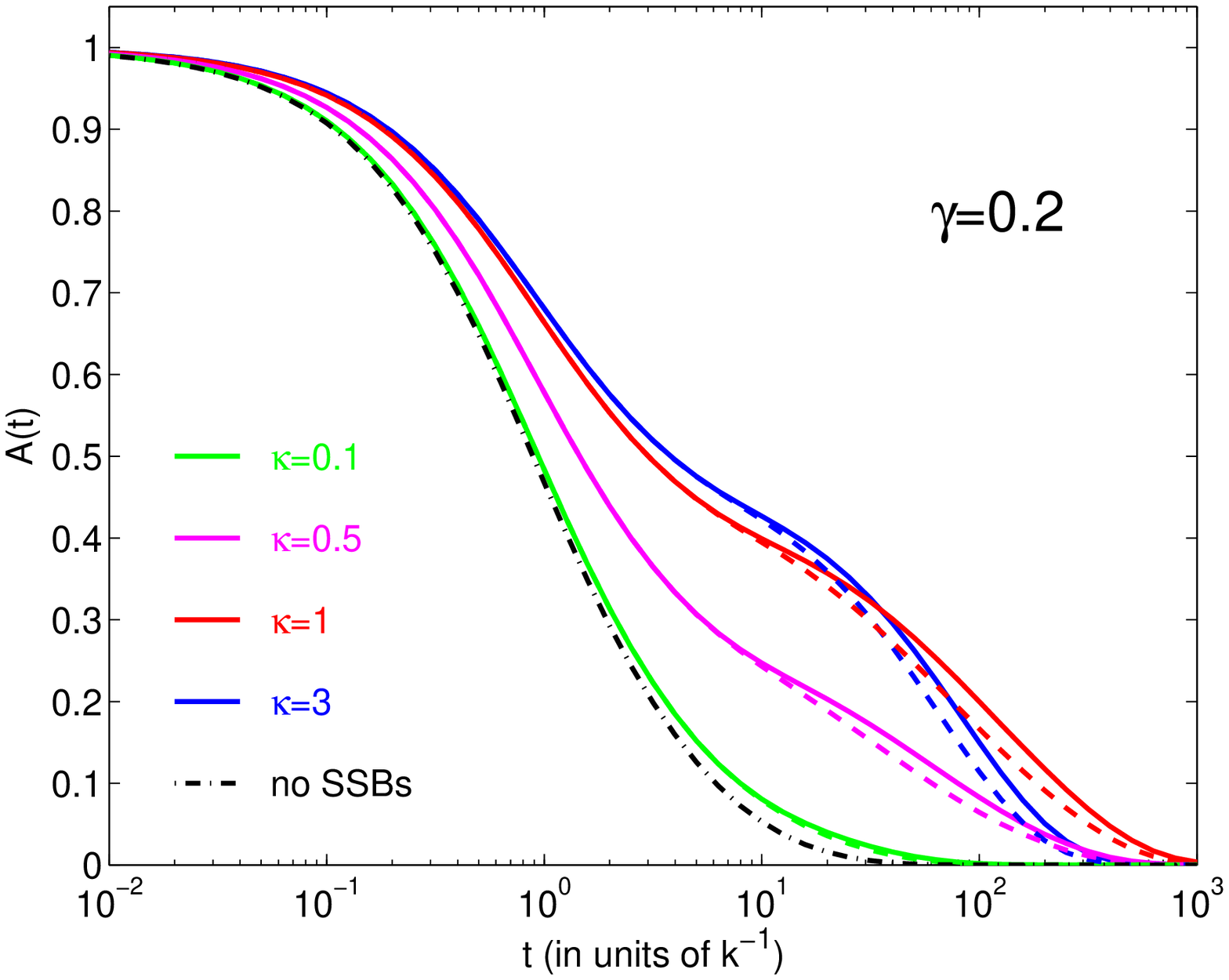}}
\end{center}
\begin{center}
\scalebox{0.42}{\epsfbox{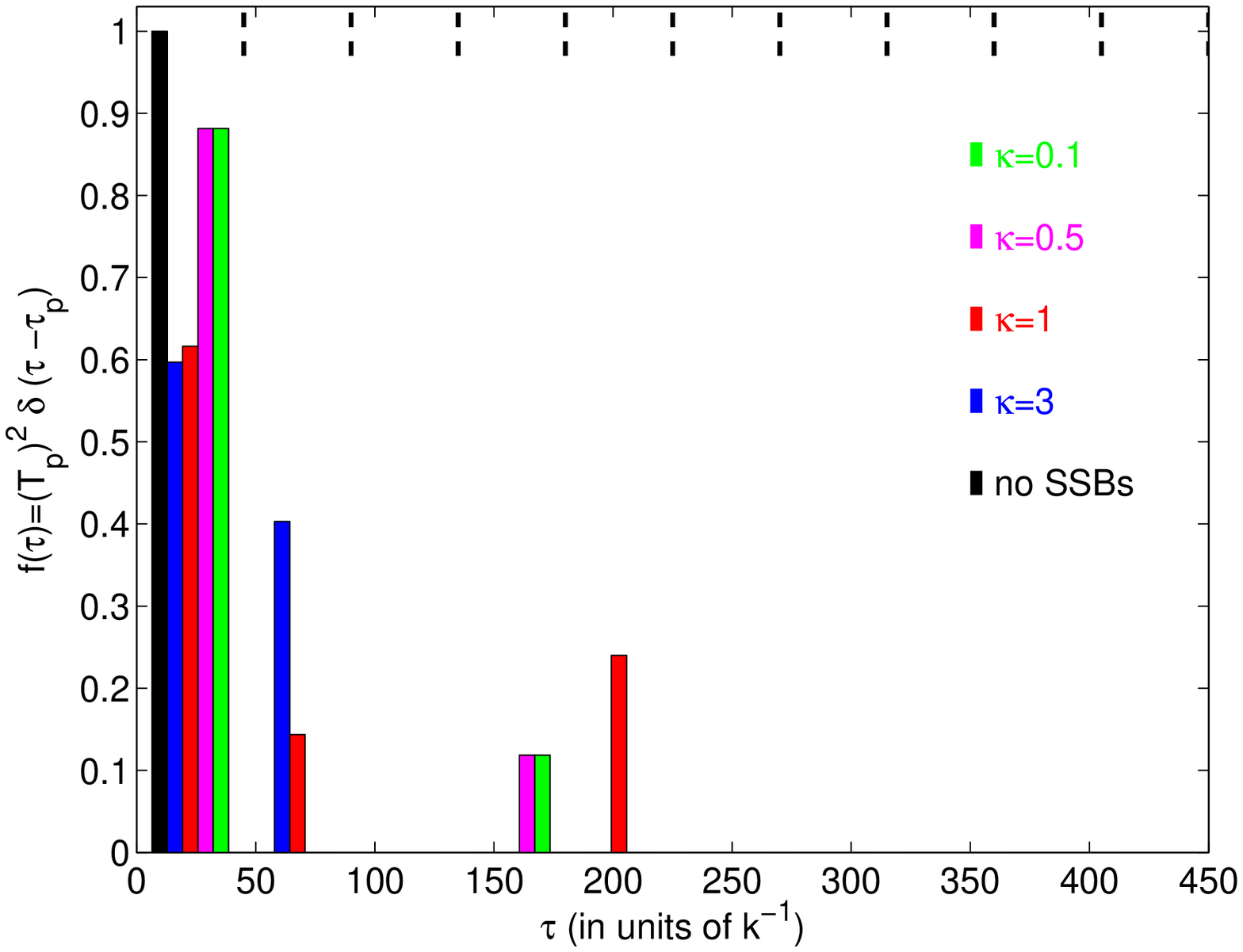}}
\scalebox{0.42}{\epsfbox{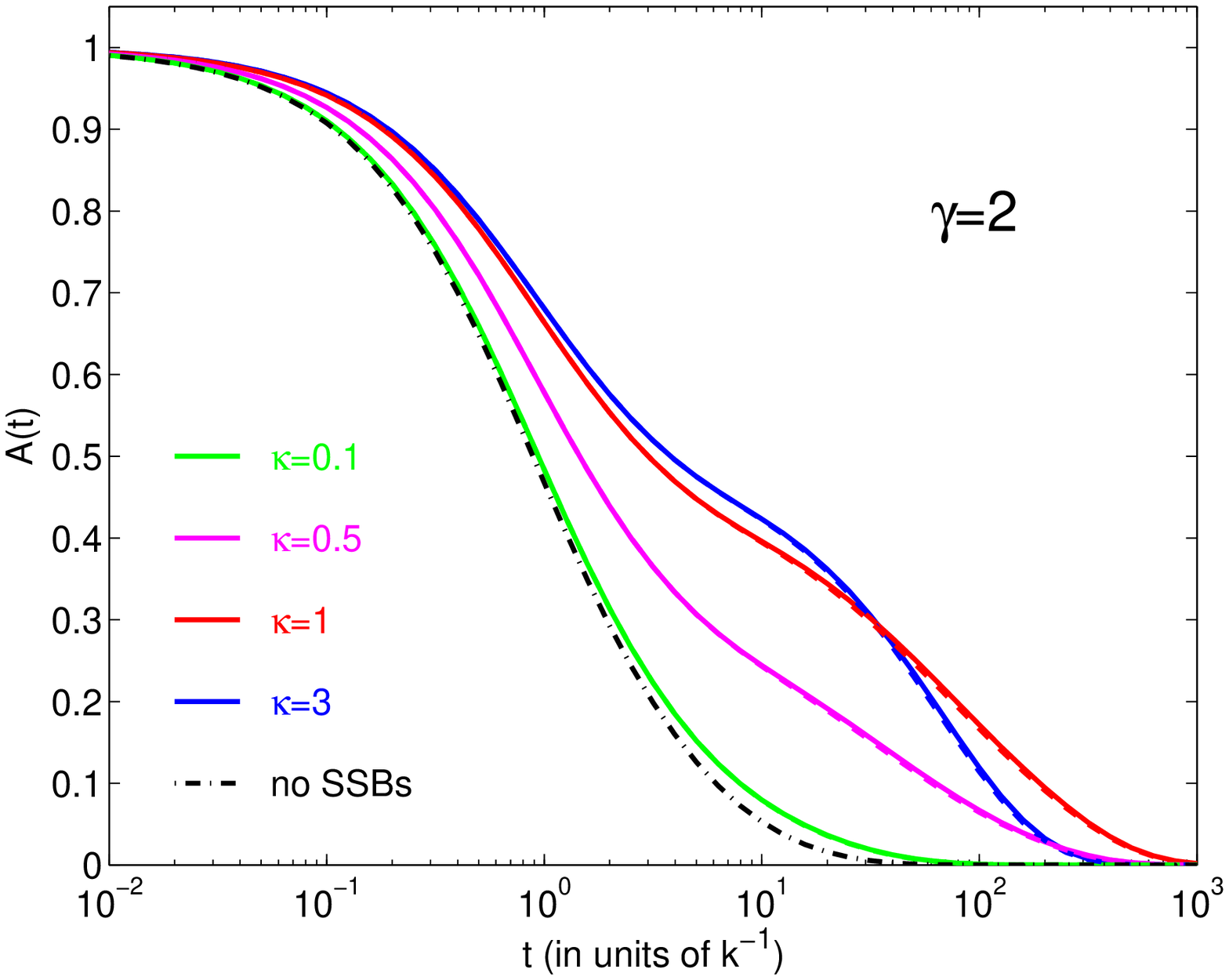}}
\end{center}
\caption{Relaxation time spectrum $f(\tau)$ and autocorrelation function
$A(t)$ for different values of the binding strength $\kappa$ and the ratio
$\gamma=q/k$ between the unbinding and the unzipping rate constants, as
indicated in the graphs. The dashed curves correspond to the approximation of
adiabatical elimination (fast binding). In the plots for the relaxation time
spectra, $f(\tau)$, the data were binned (the black dashed lines at the top of
the graphs show the bin size). We chose $u=0.6$, $\lambda=5$, $x_T=1$, and
$M=30$.
Notice that strongly binding SSBs increase the relaxation time by orders of
magnitude (logarithmic abscissa in the $A(t)$ plots).}
\label{fig:A_t}
\end{figure}

In the general case when the dynamics of SSB (un)binding and base-pair
(un)zipping occur on comparable time scales, $\gamma\sim 1$, the eigenvalue
equation \eref{eigen1} has to be solved numerically; for detailed elaboration
on the procedure, see reference \cite{ssb1}. In figure \ref{fig:A_t}, we
display some typical examples for the relaxation time spectrum and the
autocorrelation function for various values of SSB binding strength $\kappa$
and rate ratio $\gamma$.

\subsection{Discussion}

We first note that the autocorrelation function $A(t)$ in the semi-log plot
is non-exponential, corresponding to a multimodal relaxation behaviour, that
was also observed in experiment \cite{oleg}. This is reflected in the rather
broad distribution of relaxation times shown in figure \ref{fig:A_t}. In the
binned data for the relaxation times, the first bin contains the major
portion of the relaxation contributions for all cases. This describes the
relaxation of the denaturation wedge itself, i.e., the relaxation due to the 
base-pair (un)zipping process. In the plots for $A(t)$, this corresponds to
the first relaxation shoulder located at a few inverse zipping time units,
$k^{-1}$.

\begin{figure}
\begin{center}
\scalebox{0.6}{\epsfbox{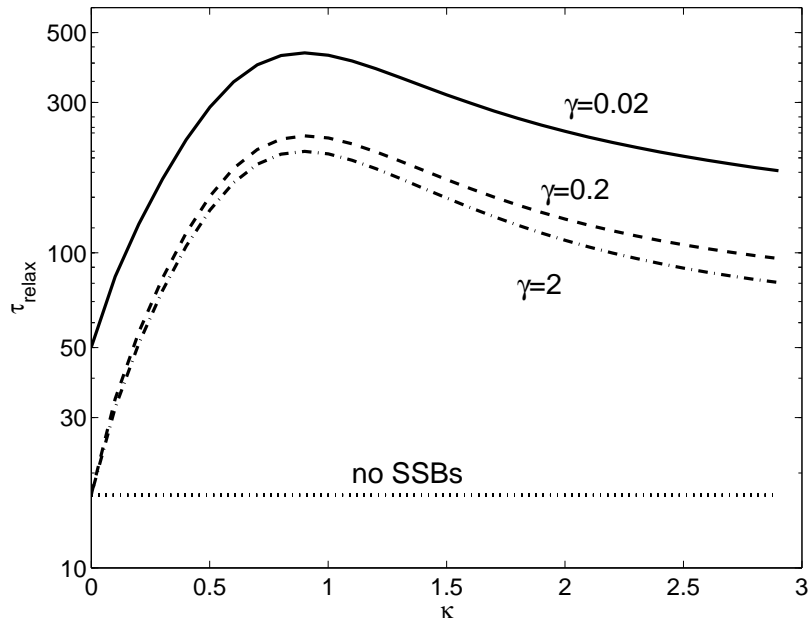}}
\end{center}
\caption{Longest relaxation time $\tau_{\mathrm{relax}}$ as a function of 
SSB binding strength $\kappa$. Note the dramatic variation of $\tau_{\mathrm{
relax}}$ (logarithmic ordinate). We chose $u=0.6$, $\lambda=5$, and $M=30$. A
maximum occurs at around $\kappa=1$.}
\label{trelax}
\end{figure}

In absence of SSBs, our coarse-grained plot of the relaxation time distribution
shows only one contribution, corresponding to only one relaxation shoulder in
the graphs for $A(t)$. With increasing binding strength $\kappa$, a second
relaxation shoulder in $A(t)$ is building up. This is due to the relaxation
of SSB (un)binding. Let us discuss the somewhat involved behaviour for
the various values of $\kappa$ in the case $\lambda=1$, for which the
occupation fraction of SSBs on the ssDNA branches of the denaturation wedge
becomes $f=\kappa/(1+\kappa)$ \cite{ambme}.
We see that $\kappa=1$ leads to an occupation
fraction $f=1/2$. In that case, we would expect the relaxation time for SSB
occupation to be the longest. Both for increasing and decreasing $\kappa$,
the fraction of vacancies and bound SSBs is decreasing, respectively, so
that the exchange between both species becomes faster, and $A(t)$ decays
quicker than for $\kappa=1$. However, for $\kappa<1$, the number of bound
SSBs is smaller and mostly a certain binding site is vacant, and the relaxation
therefore quicker than for $\kappa>1$, for which a site is mostly occupied.
Indeed, these trends can be observed in the plotted examples: The relaxation
time contribution is the longest in the case $\kappa=1$. This observation is
corroborated in figure \ref{trelax} illustrating the behaviour of the longest
relaxation time as function of binding strength $\kappa$: the maximum close
to $\kappa\approx 1$ is distinct.

In dependence of the ratio $\gamma=q/k$ of rates $q$ for SSB unbinding and $k$
for base-pair unzipping, the increase in the relative speed, $\gamma$, of the
SSB dynamics is immediately evident from the shift towards shorter $\tau$ in
the relaxation time spectrum. In the plots for $A(t)$, the results in presence
of SSBs approach the adiabatic approximation; in the case $\gamma=2$, virtually
no difference between the full result and the fast binding limit are visible in
$A(t)$.

\section{Conclusions}

We investigated by means of a master equation approach in detail the dynamics
of end-denaturation of a clamped stretch of homopolymer DNA. The transfer
rates were obtained from the partition coefficients based on the
Poland-Scheraga model of DNA denaturation. We chose the rates $\mathsf{t}^\pm$
and $\mathsf{r}^\pm$ for base-pair (un)zipping and SSB (un)binding, such that
detailed balance is fulfilled, i.e., thermal equilibrium reached for long
times. We furthermore chose these rates in a fully asymmetric
form guaranteeing that base-pair unzipping is proportional to the Boltzmann
factor for breaking the base-pair and the fundamental rate $k$. For base-pair
zipping, the rate is given by $k$ in absence of SSBs, whereas in presence of
SSBs it gets dressed by the combinatorial ratio slowing down the zipping when
SSBs are bound (essentially, by the probability that the base-pair next to the
zipping fork is vacant), and preventing closing of the wedge once it is fully
occupied.
Similarly, we chose the (un)binding rates such that the unbinding is specified
by the fundamental rate $q$ times the number of bound SSBs. Binding is biased
by the binding strength $\kappa$ and the probability of adding another SSB to
the two arches of the denaturation wedge.

Mounting a fluorophore-quencher pair at position $x_T$ on either arch of the
associated denaturation wedge, a molecular beacon is built, whose blinking
dynamics, corresponding to the open or closed state of the base-pair $x_T$,
is described by the random variable $I$. For this quantity, we obtained the
autocorrelation function $A(t)$ and the associated spectrum of relaxation
times. The quantity $A(t)$ can be measured directly in experiments (see,
e.g., references \cite{oleg,olegrev}). The predicted behaviour of $A(t)$ and
the spectrum $f(\tau)$ shows a multimodal behaviour with two pronounced
relaxation shoulders in the presence of SSBs. As our model involves physical
parameters that are known for given solvent conditions ($u$, $\kappa$) or
can be determined independently ($k$, $q$), our model becomes fully
quantitative, and can be used to devise future experimental setups.

The molecular beacon setup we propose represents the basis for an interesting
nanosensor. Constructing a small stretch of AT base-pairs and clamping them
with a few GC base-pairs at one end (compare the construction used in reference
\cite{oleg}), such a nanosensor would be of the linear size of some 20 nm. A
low concentration of such nanosensors would therefore be sufficient to probe
for presence of SSBs, salt conditions, or similar, in small probe volumes as,
for instance, encountered in gene arrays. The rather large changes in the
relaxation time spectrum invoked by the presence of SSBs (or, by varying $u$,
corresponding to temperature or salt changes, not shown here) corroborate that
this kind of sensor could actually be rather sensitive.
We note that instead of the
conventional fluorophores, that bleach rather quickly, longer-lived quantum
dots or plasmon resonant nanoparticles are now available, see, for instance, 
references \cite{dot,Schultz}.

Whereas our calculations are for a homopolymer denaturation zone, the
simplest possibility in view of designing a nanodevice, it is possible
to extend the model to a heterogeneous sequence, see reference \cite{hetero}.
However, the numerical evaluation of the corresponding master equation may
become challenging, especially, when longer denaturation zones and small
SSBs are employed. An efficient alternative for such more involved cases
is provided by stochastic simulations techniques such as the Gillespie
algorithm, as studied recently in the context of denaturation fluctuations
in DNA \cite{suman}.

\end{document}